# Infinite Physical Monkey: Do Deep Learning Methods Really Perform Better in Conformation Generation?

Haotian Zhang[1], Jintu Zhang[1], Huifeng Zhao[1], Dejun Jiang[1], Yafeng Deng[1]

[1]Hangzhou Carbonsilicon AI Technology Co., Ltd, Hangzhou 310018, Zhejiang, China

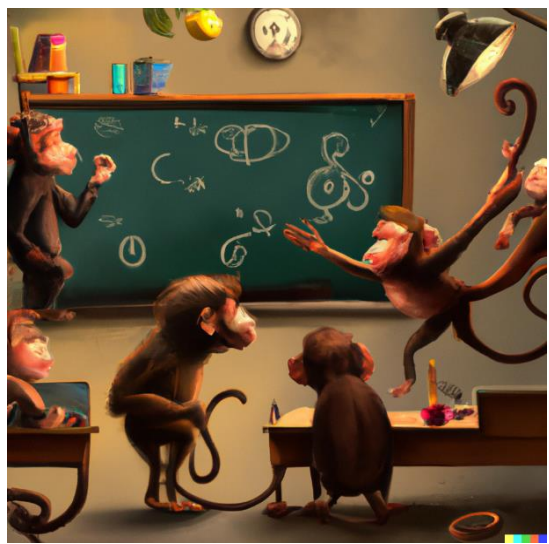

TOC: Infinite Physical Monkey. This image was created with the assistance of DALL·E 2

## Abstract

Conformation Generation is a fundamental problem in drug discovery and cheminformatics. Generally, it can be categorized into three different classes according to physical scales, i.e., micro molecule (organic), meso molecule (nano particle-like), and macro molecule (protein and nucleic acid). Organic molecule conformation generation, particularly in vacuum and protein pocket environments, is most relevant to drug design. Recently, with the development of geometric neural networks, the data-driven schemes have been successfully applied in this field, both for molecular conformation generation (in vacuum) and binding pose generation (in protein pocket). The former beats the traditional ETKDG method, while the latter achieves similar accuracy compared with the widely used molecular docking software.

Although these methods have shown promising results for real-world drug design campaigns,

some researchers have recently questioned whether deep learning (DL) methods perform better in molecular conformation generation via a "parameter-free" method. To our surprise, what they have designed is some kind analogous to the famous infinite monkey theorem, the monkeys that are even equipped with physics education. To discuss the feasibility of their proving, we constructed a real infinite stochastic monkey for molecular conformation generation, showing that even with a more stochastic sampler for geometry generation, the coverage of the benchmark QM-computed conformations are higher than those of most DL-based methods. By extending their physical monkey algorithm for binding pose prediction (with 2000 random samples), we also discover that the successful docking rate also achieves near-best performance among existing DL-based docking models. Thus, though their conclusions are right, their proof process needs more concern.

In addition to evaluating the rationality of their algorithms and conclusions, we dig into the inspirations of infinite physical monkeys. We find that, for docking pose generation, DL-based models truly learn the interaction rules between residues and ligands, and discover an inductive bias hidden in the training of the pocket-given docking problem.

The code of the proposed algorithm could be found at:

https://github.com/HaotianZhangAI4Science/infinite-physical-monkey.

## Introduction

Binding pose generation is a fundamental and significant problem in drug discovery and cheminformatics[1,2], which is aimed at obtaining the spatial and orientational relationship between protein pockets and drug candidates. Once the binding pose is determined, it can be utilized for structure-based drug design (SBDD) and ligand-based drug design (LBDD) based on protein-ligand energy profiles. In particular, high-throughput virtual screening[3], the most popular SBDD protocol, requires the binding energy assessment for every molecule in the screened compound library, in which the scoring function highly relies on the plausibility of binding poses.

Numerous algorithms have been developed to predict binding poses, with some acknowledged as standard tools in virtual screening protocol[4]. Docking pose generation algorithms could be classified into systematic search, heuristic search, and deterministic search based on the sampling strategy. The systematic search represented by Glide[5] samples all degrees of freedom, but the sampling complexity grows exponentially with the number of rotatable bonds, contributing to the curse of dimensionality. Therefore, the search space is first filtered according to empirical statistical rules in practical applications. Heuristic search involves performing random transformation to the ligand conformations, which are then accepted or rejected based on the evaluation function, iterated until convergence. In particular, AutoDock[6] and LeDock[7] are two kinds of this class that use the genetic algorithm and simulated annealing algorithm, respectively Deterministic search, which involves using molecular dynamics as the search engine, explores the conformational space with a greedy strategy, moving step by step toward lower energy regions. However, since deterministic search collapses to the same final state with the same initial state and the same dynamics parameters, it is less common than the first two. CDOCKER[8] is an example of deterministic software based on the CHARMM[9] molecular dynamics simulation. Although these molecular docking algorithms have played a significant role in drug discovery campaigns over the past few decades, the intrinsic complexity of protein-ligand interactions consisting of entropy and solvation effects[10,11] make the binding pose generation far from solved. Accurate prediction of binding poses remains a key problem in computer-aided drug design (CADD).

In recent years, the data-driven scheme has been utilized for tackling this problem and has shown powerful potential in real-world scenarios. DeepDock[12] follows the heuristic search ideas, training a scoring function to accept and reject change transformations of conformations. EquiBind[13] proposes a SE(3)-equivariant framework to embed the physical constraints when predicting Cartesian coordinates directly. TankBind[14] generates extensive 2D information, i.e., interatomic distances inside ligands and between ligand atoms and protein atoms, to recover the 3D conformation. However, since it's not a one-to-one mapping, the recovering process would compromise the plausibility of binding poses. Thus physical refinement is often required to polish these conformations[15]. A more physical model, DiffDock[16], has been designed to operate on the internal coordinate space with the popular and powerful diffusion model, achieving the state-of-the-art (SOTA) performance among all the deep learning (DL)-based methods. The foundation of these rapid developments is from the molecular conformation generation methods, like GeoDiff[17] for directly generating Cartesian coordinates, CGCF[18] and SDEGen[15] for constructing

3D geometries from extensive 2D distances, and torsional diffusion[19] for operating search in torsional angle space.

Table 1. The evaluation contents in different works

| Methods | COV/MAT | Dist. Distri. MMD | Thermo. Prop. | Others |
|---|---|---|---|---|
| GraphDG |  | ☑ | ☑ |  |
| CVGAE | ☑ |  |  | Diversity |
| ConfGF | ☑ | ☑ | ☑ |  |
| CGCF | ☑ | ☑ |  |  |
| ConfVAE | ☑ | ☑ |  | Diversity |
| GeoMol | ☑ |  |  |  |
| GeoDiff | ☑ |  | ☑ |  |
| SDEGen | ☑ | ☑ | ☑ | Energy Prof. Crystal Comp. |
| Tor. Diff. | ☑ |  |  |  |
| DMCG | ☑ |  | ☑ | Docking |
| DGSM | ☑ |  | ☑ |  |

Recently, there is a work[20] showing the DL-based model was defeated by their curated parameter-free method. This method sampled ~2000 conformations from RDKit by rotating the dihedral angles in 1/6 of them, performing force field optimization in 1/6 of them, and clustering the resulting conformations based on the 3D coordinates. This constructed baseline achieves comparable results to the SOTA method in terms of the COV and MAT metrics, leading to questions about the rationality of the benchmark in the DL-based approach. Two points need to be noted: first, for the molecular conformation generation problem, COV and MAT are only a part of the assessment metrics, and other evaluation approaches have been used in the field. "*By revising the MCG setting in many DL-based methods, we suggest the community rethink the benchmark in the current MCG, and focus on the end applications in various MCG-related downstream applications in the future*"[21], what they have suggested have long been adopted among most of the molecular conformation generation works. For example, in GraphDG[22], ConfGF[23], and GeoDiff[17], the difference between the thermodynamic properties of the generated conformations and quantum-computed conformations was compared; in DMCG[24], it is demonstrated that the docking success rate was improved when the model-generated conformations were taken for downstream docking tasks; in SDEGen[15], not only the generated conformations were compared with the crystal conformations, but the free energy profile was also plotted by molecular dynamics to qualitatively compare the coverage area of the model-generated conformations with that of the RDKit-generated conformations. **Table 1** lists the evaluation methods involved in each model in the field of molecular conformation generation, emphasizing that there are a variety of metrics adopted for evaluating conformation generation models. Second, the parameter-free method they constructed beat other machine learning models because of the oversampling, which coincides with the infinite monkey theorem[25], the monkeys were even equipped with physical knowledge. This prompted the construction of a true infinite stochastic monkey algorithm in this paper to demonstrate that even this more random method can achieve the reported accuracy. Compared to their infinite physical monkey, the infinite stochastic monkey goes further away from the dependence on the parameters of the ETKDG algorithm, and the idea of this approach is as follows: we model the chemical bonds using a resonant oscillator model, under which the bond length distribution can be well approximated by a Gaussian distribution. The mean and standard deviation of the bond lengths for different bond types (e.g., carbon-carbon single bond, carbon-oxygen double bond) in the data set are counted. During the generation procedure, we resample each chemical bond length and then reconstruct the three-dimensional geometries from the two-dimensional bond lengths. The experiments demonstrate that our method also achieves almost as close to their proposed RDKit+Clustering algorithm, indicating that the comparison in the previous is unfair. At the very least, they should also sample 2000 conformations for DL methods followed by clustering.

We also extended the infinite monkey algorithm for the binding conformation prediction problem and constructed the Scoring baseline considering pocket information and the UFF baseline model based on force field optimization. Our approach highlights the importance of large sampling size in achieving high success rates in molecular docking, challenging the use of RDKit+Clustering as a fair baseline model. Moreover, we show that the infinite physical monkey can serve as a stochastic baseline, shedding light on the binding conformation prediction problem. **Table 6** presents the docking success rates of Uni-Dock, two traditional docking methods, two DL methods given pockets, and our constructed baselines. Our results suggest that machine learning models have the potential to surpass traditional models in terms of the docking success rate for a given pocket. Additionally, any DL-based docking model must outperform the infinite physical monkey under the same conditions to validate its effectiveness. We also identify the possible inductive bias of current pocket-based docking models and propose straightforward training methods to mitigate this bias.

## Methods

**Infinite Stochastic Monkey (Molecular Conformation Generation)**

As we mentioned above, the RDkit+Clustering algorithm is essentially a physical version of the infinite monkey theorem, like shooting a target with a machine gun and then clustering the scores. With the same sample size, COV/MAT can reflect the conformational quality to some extent, not to mention that in the field of molecular conformation generation, there are at least distance distribution and thermodynamic calculation experiments to further verify the quality of conformation ensemble.

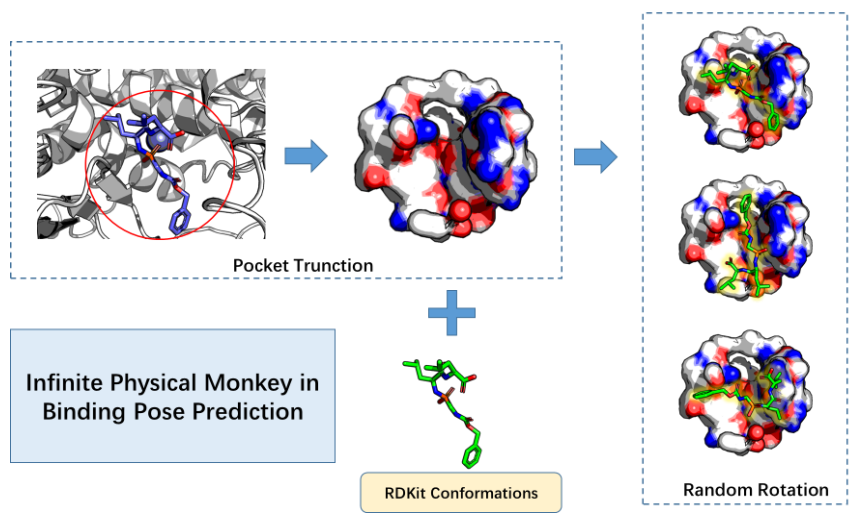

Figure1. The workflow of Infinite Physical Monkey algorithm in Binding Pose Prediction

To further demonstrate that even a random algorithm can achieve the SOTA performance on COV/MAT with a large sample size, we devise a truly parameter-free random method, named as Infinite Stochastic Monkey. As shown in **Figure 1**, this algorithm goes as follows: 1) a random sampler is used to generate bond lengths by assuming a Gaussian distribution based on the resonator model. The resulting pseudo-distance matrix is then used to reconstruct the three-dimensional conformation of the molecule. 2) an ETKDG-constrained dihedral random sampler is applied to generate a series of conformations that satisfy the ETKDG algorithm distribution for bond lengths and angles but have a uniform random dihedral distribution. 3) a force-field sampler is utilized to optimize the conformations obtained from the previous first step using the MMFF force field optimization. Compared with the Infinite Physical Monkey (RDKit+Clustering), our method further gets rid of the reliance on prior knowledge, and the sampling speed increases ~10 times or so. The sample size follows the original survey[21] as $N_e$:

$$N_e = \min(20 N_{ref}, 2000)$$

where $N_{ref}$ is the number of the favored conformations at the quantum chemistry accuracy in the dataset, the number of ETKDG-constrained dihedral random conformations and force-field optimized conformations is 1/4 of randoms. Similarly, the K-means algorithm is adopted to cluster twice as many conformations as $N_{ref}$, and the center of each cluster is taken as pseudo-conformations for evaluation.

**Infinite Physical Monkey (Binding Pose Prediction)**

Except for the high performance in the molecular conformation generation, similar results could also be reached by the infinite physical monkey algorithm in the binding pose generation problem. Among drug design applications, the most commonly used scheme is predicting the binding conformations under pocket given conditions. Based on the assumption that the pocket is given, the infinite physical monkey algorithm is as follows: 1) generate the molecular conformation using the ETKDG algorithm, where the conformation is generated without considering any protein environment; 2) place the molecule onto the geometric midpoint of the pocket and then perform a random rotation centered on the geometric midpoint to obtain the final infinite physics monkey version of the docking conformation. Moreover, we propose a scoring baseline model that takes into account the chemical environment inside the pocket, i.e., after obtaining the docking conformation of the infinite physical monkey, the binding energy of the conformation inside the pocket is evaluated by Vina[26], and the molecules are ranked in the order of the binding energies from lowest to highest. In addition to this post-processing approach, we also propose a force field-based baseline for binding conformation prediction: perform force field optimization on molecules using the infinite physical monkey conformation as a starting point with residues fixed in the pocket. With the infinite physical monkey algorithm built on the molecular conformation generation task and binding pose prediction task, it can be concluded that a large sampling size results in a sustained improvement of the metrics associated with RMSD.

Thus, we propose suggestions to the method comparison in this discussed paper[21], at least, when they use infinity monkeys of RDKit, they should also make other machine learning methods infinity monkeys. Although we partially agree that COV/MAT is not a completely reasonable metric, there is no perfect metric in the world, and if we compare a method to an elephant, then each metric is like a blind person who touches the elephant and uses different metrics for evaluation in order to understand the overall shape of this elephant as much as possible. Unfortunately, they only asked a blind person who was touching the elephant's leg what the elephant really looked like, ignoring the opinions of other blind people.

**Method comparsion in binding pose prediction given pockets**

We agree Yu et al. [20] that DL model should focus on the local docking scenario, which predicts the binding conformations inside the protein pockets rather than within the whole protein. However, there are still some DL-based docking models developed for fair comparison. In the last experiment, we discuss the results of the pocket-aware docking methods, from traditional approaches, such as AutoDock GPU, Glide SP, and Uni-Dock, to the deep-learning models, such as LigPose[27] and TankBind[14]. Besides, three baselines are also present for demonstrating that the DL-based models truly model the interaction between the ligands and pockets.

# Results and discussions

**Experiment 1: Performances of Infinite Stochastic Monkey in Molecular Conformation Generations.**

Although our Infinite Stochastic Monkey algorithm is much more stochastic than the RDkit+Clustering methodology, it still achieves the SOTA performance on COV/MAT. Such results indicate that under the large sampling number limit (e.g., 2000 conformations), pure stochastic algorithms should give out the limiting COV/MAT scores. And this is easy to understand: when performing conformation generations with a large sampling number, it is equal to uniformly sampling the molecules' conformation spaces. This is just like randomly sanding infinite monkeys (trial conformations) to the molecules' potential surfaces, and then analyzing if the monkeys could fully occupy the surfaces. And the results must be deterministic: the sampled conformation space would naturally cover all low-energy conformations, including the reference conformations. In the case of RDkit, the monkeys are further driven by the physical rule and are only sent to the regions with relatively low potential energies, which is just the so-called enhanced sampling methodology. As a result, under a large but not infinite sampling number, the monkeys with physical knowledge in their mind perform a little bit well than our pure stochastic monkeys, as can be seen in **Table 2.**

Table 2. COV/MAT metric on GEOM-QM9 and GEOM-Drugs.

| Method | QM9 | | | | Drugs | | | |
|---|---|---|---|---|---|---|---|---|
| | COV(↑, %) | | MAT(↓, Å) | | COV(↑, %) | | MAT(↓, Å) | |
| | Mean | Median | Mean | Median | Mean | Median | Mean | Median |
| RDKit | 81.82 | 85.98 | 0.3027 | 0.2564 | 70.47 | 77.08 | 1.2069 | 1.1080 |
| CVGAE | 0.09 | 0.00 | 1.6713 | 1.6088 | 0.00 | 0.00 | 3.0702 | 2.9937 |
| GraphDG | 73.33 | 84.21 | 0.4245 | 0.3973 | 8.27 | 0.00 | 1.9722 | 1.9845 |
| CGCF | 83.48 | 86.70 | 0.2984 | 0.2694 | 72.41 | 74.09 | 1.1198 | 1.1017 |
| ConfVAE | 80.42 | 85.31 | 0.4066 | 0.3891 | 53.14 | 53.98 | 1.2391 | 1.2447 |
| ConfGF | 88.49 | 94.13 | 0.2673 | 0.2685 | 62.15 | 70.93 | 1.1629 | 1.1596 |
| GeoMol | 71.26 | 72.00 | 0.3731 | 0.3731 | 67.16 | 71.71 | 1.0875 | 1.0586 |
| DGSM | 91.49 | 95.92 | 0.2139 | 0.2137 | 78.73 | 94.39 | 1.0154 | 0.9980 |
| SDEGen | 92.14 | 99.55 | 0.2035 | 0.2002 | 92.00 | 98.15 | 0.7892 | 0.7665 |
| GeoDiff | 92.65 | 95.75 | 0.2016 | 0.2006 | 88.45 | 97.09 | 0.8651 | 0.8598 |
| DMCG | 94.98 | 98.47 | 0.2365 | 0.2312 | 91.27 | 100.00 | 0.8287 | 0.7908 |
| Inf. Phy. Monkey[#] | 97.65 | 100.00 | 0.1902 | 0.1818 | 87.93 | 100.00 | 0.8086 | 0.7838 |
| Inf.Sto.Monkey | 86.36 | 91.73 | 0.3569 | 0.3560 | 86.40 | 98.27 | 0.9662 | 0.9642 |

[#]Inf. Phy. Monkey (RDKit+Clustering)

On the other hand, DL methods are designed to generate low-energy conformations with a small sampling number. Thus, it is unfair to criticize the DL method for not outperforming the limiting case of the RDkit+Clustering methodology, especially when the sampling number of the DL methods are far smaller than the stochastic algorithms. What's more, as we mentioned above, since the COV/MAT scores have their own limitations, the SMCG community no longer invokes them as the only benchmark standard. So it is also meaningless to only use them as the criteria when benchmarking the performances of SMCG methods.

**Experiment 2: Performances of Infinite Physical Monkey in Binding Pose Prediction.**

Table 3. Hit rate of Infinite Physical Monkey baseline

| Num | Threshold (Å) |
|---|---|

| Confs. | 1.0 | 1.5 | 2.0 | 2.5 | 3.0 | 3.5 | 4.0 | 4.5 | 5.0 |
|---|---|---|---|---|---|---|---|---|---|
| 2000 | 28.71 | 54.41 | 74.84 | 87.62 | 94.01 | 97.04 | 98.86 | 99.71 | 99.81 |
| 1000 | 27.28 | 52.43 | 73.13 | 86.77 | 93.84 | 96.73 | 98.86 | 99.71 | 99.81 |
| 500 | 21.96 | 47.99 | 70.27 | 84.88 | 92.79 | 96.18 | 98.65 | 99.71 | 99.81 |
| 100 | 8.40 | 29.81 | 55.80 | 72.19 | 86.52 | 93.33 | 96.37 | 99.32 | 99.81 |
| 50 | 4.44 | 22.31 | 45.16 | 61.93 | 75.83 | 88.17 | 93.99 | 98.17 | 98.65 |
| 20 | 0.81 | 11.42 | 32.23 | 49.42 | 62.53 | 75.25 | 86.32 | 93.73 | 97.45 |
| 10 | 0.41 | 5.31 | 16.77 | 35.17 | 49.91 | 64.67 | 74.92 | 85.59 | 92.56 |
| 5 | 0.41 | 3.27 | **9.41** | 18.42 | 31.94 | 50.79 | 58.17 | 72.10 | 80.30 |
| 1 | 0 | 0 | **1.63** | 5.73 | 7.78 | 19.26 | 28.28 | 40.16 | 48.77 |

To further illustrate that numerous samples could cover most of the energy profile, we curated the Infinite Physical Monkey algorithm for a well-acknowledged problem, docking pose generation. We aligned the molecules whose conformations are generated by RDKit to the center of the protein pocket followed by a random rotation. **Table 3** shows the performance of the constructed Infinite Physical Monkey under the different number of conformations and thresholds. The result of the 2A-hit rate under 2,000 samples, 74.84%, is close to the SOTA performance, outperforming most traditional docking methods. But you will never take these resulted conformations to assess the binding energy for virtual screening in real-world drug development, which is the same discussion with the previous Infinite Physical Monkey in molecular conformation generation, also namely the RDKit+Clustering.

Table 4. Hit rate of Random Scoring baseline

| Num Confs. | Threshold (Å) | | | | | | | | |
|---|---|---|---|---|---|---|---|---|---|
| | 1.0 | 1.5 | 2.0 | 2.5 | 3.0 | 3.5 | 4.0 | 4.5 | 5.0 |
| 2000 | 22.94 | 57.64 | 80.58 | 89.41 | 95.88 | 97.05 | 98.82 | 99.41 | 99.41 |
| 1000 | 22.94 | 57.64 | 80.58 | 89.41 | 95.88 | 97.05 | 98.82 | 99.41 | 99.41 |
| 500 | 14.70 | 45.29 | 72.94 | 85.88 | 94.70 | 96.47 | 98.82 | 99.41 | 99.41 |
| 100 | 7.05 | 24.11 | 51.76 | 68.82 | 83.52 | 93.52 | 96.47 | 97.64 | 98.82 |
| 50 | 2.35 | 13.52 | 36.47 | 56.47 | 73.52 | 87.05 | 94.11 | 95.88 | 98.23 |
| 20 | 1.17 | 7.64 | 24.11 | 44.11 | 58.82 | 77.05 | 88.82 | 92.35 | 95.29 |
| 10 | 1.17 | 3.52 | 17.64 | 33.52 | 47.05 | 61.76 | 77.64 | 84.11 | 91.76 |
| 5 | 0.58 | 1.76 | **12.35** | 22.94 | 31.17 | 44.70 | 58.82 | 75.88 | 83.52 |
| 1 | 0 | 0 | **1.76** | 4.70 | 7.64 | 13.52 | 25.88 | 35.88 | 47.05 |

Table 5. Hit rate of UFF baseline

| Num | Threshold (Å) |
|---|---|

| Confs. | 1.0 | 1.5 | 2.0 | 2.5 | 3.0 | 3.5 | 4.0 | 4.5 | 5.0 |
|---|---|---|---|---|---|---|---|---|---|
| 2000 | 69.23 | 83.58 | 91.28 | 96.92 | 99.46 | 99.48 | 99.48 | 99.48 | 100.0 |
| 1000 | 69.23 | 83.58 | 91.28 | 96.92 | 98.46 | 98.48 | 98.48 | 98.48 | 100.0 |
| 500 | 63.07 | 79.84 | 88.71 | 94.87 | 96.41 | 98.46 | 99.48 | 99.48 | 100.0 |
| 100 | 47.17 | 62.54 | 77.43 | 83.58 | 87.17 | 92.30 | 97.43 | 97.94 | 98.46 |
| 50 | 34.87 | 51.79 | 66.66 | 73.33 | 78.97 | 86.15 | 90.25 | 95.89 | 96.92 |
| 20 | 24.10 | 38.97 | 49.23 | 60.51 | 67.69 | 76.41 | 80.51 | 86.66 | 89.74 |
| 10 | 16.41 | 27.69 | 38.97 | 45.64 | 54.35 | 63.07 | 68.20 | 72.3 | 80.0 |
| 5 | 11.79 | 18.46 | **26.66** | 32.30 | 38.46 | 47.69 | 54.35 | 60.0 | 66.15 |
| 1 | 4.10 | 6.15 | **9.23** | 12.30 | 15.38 | 21.53 | 26.15 | 31.79 | 40.0 |

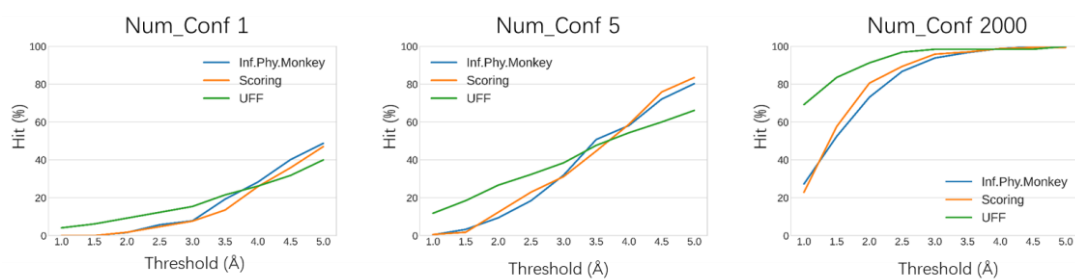

**Figure2.** The Hit rate of three baseline methods at different thresholds

Now we could further investigate the revelations taken by our monkeys. We calculated the binding energy between the randomly generated ligand conformations and the protein pocket. Then we sorted the RMSD values between the generated ligand conformations and the crystal ligand conformation. The results are presented in **Table 4.** We found that after the sorting, the top-1 and top-5 2A hit rates are raised by a little compared with the original Infinite Physical Monkey algorithm. This is because, during the sorting, we put the more reasonable binding conformations (the conformations with higher binding scores) at the top of the conformation list. But after all, these conformations were generated from the random rotating, thus most of the conformations are not superior in binding energies. Some of them will even overlap with the atoms of the binding pocket. This experiment further explained the randomness of our Infinite Physical Monkey algorithm. But under the same sampling number, the docking success rate is still a valid benchmarking score.

Based on the conformations generated by the Infinite Physical Monkey algorithm, we further tested if force-field-based geometry optimization could improve the docking success rates. The geometry optimization process could move the conformations to the nearest local potential energy minima. Thus accuracies of the optimized geometries depend on the initial guess structures and the accuracy of the chosen force field. And in the current experiment, we invoked the UFF force field. As shown in **Table 6**, compared with random conformations, the top-1 and top-5 2A hit rates of the optimized conformations raised to 9.23% and 26.66%. And we plotted the hit rates of three selected methods under a sampling number of 1, 5, and 2000 in **Figure 2.** From this figure, we could summarize at least two conclusions: first, once a DL model could outperform the baseline version of our Infinite Physical Monkey algorithm, it must have learned the distribution of the molecules inside the protein pocket. And we found that each currently developed models are better than the stochastic method, which proves the effectiveness of these models. Second, when developing deep-learning-based binding conformation prediction models, we should pay more attention to the truncation method of the binding pocket. In most training tasks, we tend to select the binding pocket according to the center of the ligand. As a result, the training sets we constructed usually only contain one kind of conformation, in which the center of the ligand is the center of the binding pocket. The models trained with such training sets would naturally tend to place the ligand at the center of the given pocket, even if the true binding pocket does not locate at the center of the given residues. A simple methodology to avoid such inductive bias is adding Gaussian random noises to the center position of the ligand when extracting the binding pocket:

$$r_c = r_l + \mathcal{N}(r_l, 1\text{Å})$$

$$\{\text{atom}_j | \|\text{atom}_j - r_c\|^2 < \delta\}$$

In the above equation, $r_l$ is the Cartesian coordinates of the ligand center, $r_c$ is the Cartesian coordinates of the pocket center, $\{\text{atom}_j\}$ is the j-th pocket atom, and $\delta$ is a given threshold.

**Experiment 3: The Comparisons of Different DL Docking Models.**

Table 6. Success Rate of each method and its belonging class

| Method | Success Rate | Class |
| --- | --- | --- |
| Uni-Dock | 51.1% | DP technology |
| AutoDock GPU | 19.70% | Traditional |
| Glide SP | 66.8% | Traditional |
| TankBind | 24.2% | Deep Learning |
| LigPose | 74.7% | Deep Learning |
| Inf. Phy. Monkey | 1.63% | Random |
| Scoring | 1.76% | Random |
| UFF | 9.23% | Force Field |
| Inf.Phy.Monkey-Top5 | 9.41% | Random |
| Scoring-Top5 | 12.35% | Force Field |

In **Table 6** we compared the binding-pocket-specific methods (which require the users to specify the binding pocket before docking) with the deep-learning-based local docking models. For the sake of fairness, two traditional methods (AutoDock, Glide SP), two DL models, and three baselines contrasted by us were selected for the comparison. The comparison standard is the top-1 2 Å hit rate of the methods. As can be seen, the traditional methods could reach a maximum accuracy of 66.8% (Gilde SP), while the DL methods could reach a SOTA accuracy of 74.7% (LigPose). Such a result states that DL methods have the ability to outperform the traditional methods in the binding conformation prediction tasks when binding pockets are given. Another deep-learning-based method, which is the TankBind model, although only achieves an accuracy of 24.2%, still performs much better than our three baselines. As we figured out, once a DL model could outperform the pure stochastic or scoring-based Infinite Physical Monkey method, the model must have learned the interactions between the pocket and the ligands. In conclusion, DL models could be superior to traditional methods in binding conformation generations. And for both classes of methods, the ultimate way to benchmark the effectiveness is by performing the subsequent biological activity experiments.

## Conclusions

In this paper, we systematically analyze the plausibility of comparsion made by Infinite Physical Monkey algorithm in molecular conformation generation and binding pose prediction. The conclusions are as follows:

1) Although COV/MAT has their own limitations, but the reason why RDKIt+Clustering outperforms other deel learning-based model is primarily due to the larger samling size. And in the field of molecular generation, there are other physics-related evaluation metrics..

2) The Infinite Physical Monkey provides a stochasic baseline for evaluating the feasibility of DL-approch in binding pose predition.

3) Our fair comparison demonstrates that DL approaches have the potential to enhance binding pose prediction. Moreover, this comparison reveals an inductive bias that is hidden in the pocket truncation process during data processing.

In summary, our analysis suggests that the use of physics-based metrics and the consideration of sample size are important factors in evaluating the performance of molecular conformation generation and binding pose prediction methods. Furthermore, our results highlight the potential of DL approaches and the importance of considering inductive biases in the data processing phase.

## Data and Code Availability

The data and source code of this study is freely available at GitHub ([https://github.com/github.com/HaotianZhangAI4Science/infinite-physical-monkey](https://github.com/github.com/HaotianZhangAI4Science/infinite-physical-monkey) ) to allow replication of the results.

## Acknowledgments

We thank Jianfei Song, Hui Shi and many colleague in Carbon Silicon AI for their great help in this project.